# Prediction of the Reactivity of Argon with Xenon under High Pressure

Xiao Z. Yan,[1, 2] Yang M. Chen,[1, 2] and Hua Y. Geng,[1, *]

[1]*National Key Laboratory of Shock Wave and Detonation Physics, Institute of Fluid Physics, CAEP; P.O. Box 919-102, Mianyang, Sichuan, People's Republic of China, 621900*

[2]*School of Science, Jiangxi University of Science and Technology, Ganzhou, Jiangxi, People's Republic of China, 341000*

**ABSTRACT**

Pressure significantly modifies the microscopic interactions in condense phase, leading to new patterns of bonding and unconventional chemistry. Both argon and xenon possess closed-shell electronic structures, which renders them chemically unreactive. Using unbiased structure searching techniques combined with first-principles calculations, we demonstrate the reaction of argon with xenon at a pressure as low as 1.1 GPa, producing a novel van der Waals compound $XeAr_2$, which crystallizes in the $MgCu_2$-type Laves phase structure. Due to the pressure-induced delocalization of the electrons in outermost shells, the covalent Xe-Xe and Xe-Ar bonds have been detected which lead $XeAr_2$ to be unexpectedly stable without any phase transition or decomposition at least up to 500 GPa.

[*]To whom correspondence should be addressed. Email: s102genghy@caep.cn.

## 1. INTRODUCTION

Due to the stable closed-shell electronic configuration, noble gas (NG) elements (e.g. He, Ne, Ar, Kr and Xe) were historically believed to be chemically unreactive. However, Pauling[1] predicted that Xe and Kr may bond to electronegative atoms, which was proved experimentally by Bartlett[2] with the synthesis of first NG compound $XePtF_6$. This seminal discovery leaded to a recognized renaissance in NG chemistry in the past few decades[3-5].

Under ambient conditions, the heavier NG element Xe, and to some extent, Kr and Ar, have been known to be oxidized by halogen and oxygen, forming halide and oxide compounds[6-8]. Under high pressures, the reactivity of NGs was drastically altered.

On one hand, application of high pressure makes NGs easier to be oxidized; for example, NGs can be oxidized not only by fluorine[9-12] and oxygen[13-16], but also nitrogen[17] and sulphur[18], and even metal elements such as iron and nickel[19]. Theoretical predictions revealed that Xe reacts with S and N at 146 GPa[17] and 191 GPa[18], respectively. Direct reactions of Xe with Fe/Ni were predicted to occur under the pressures of Earth's inner core[19]. A subsequent experimental study reported the synthesis of Xe-Ni compound at a pressure round 150 GPa[20]. Of special attention, the Fe/Ni atoms in the Xe-Fe/Ni compounds play a role of anions instead of cations that behave as usual metals. In addition, Xe has also been observed to react with water ice at pressures above 50 GPa[21]. In these NG compounds, Xe bonds to other elements chemically by sharing its closed shell electrons. On the contrary, NG elements can also become oxidants and gain electrons from alkali and alkaline earth metals such as Li[22], Cs[23] and Mg[24].

On the other hand, NGs also form van der Waals (vdW) compounds under lower pressures wherein the NG atoms do not lose or gain electrons. For instance, the Laves phases compounds $NeHe_2$[25], $ArHe_2$[26], $Ar(H_2)_2$[27], $Xe(N_2)_2$[28,29] and $Xe(O_2)_2$[30] can be synthesized at pressures of a few gigapascals. The stabilities of such compounds can be explained in terms of packing rules analogous to binary crystals of hard spherelike particles for intermetallic compounds. Other classes of vdW compounds have also been discovered. $(N_2)_6Ne_7$[31], the structure can be viewed as a clathrate with the centers of the $N_2$ molecules forming distorted dodecahedron cages, each enclosing 14 Ne atoms. $XeHe_2$[32] stabilizes at 12 GPa, adopting a hexagonal $AlB_2$-type structure. Some other stoichiometries such as the Xe-$H_2$[33-35], Xe-$H_2O$[36] and He-$H_2O$[37 38] systems have also been observed. The origin of the stability for these compounds was not well understood.

In a few cases, unexpected chemistry in Xe-F compounds stabilized by covalent Xe-Xe bonding[9] and in Na$_2$He compound stabilized by long-range Coulomb interactions[39] have also been reported. It is obvious that the bonding of NGs in compounds under high pressures exhibits strong uncertainty, which provides a very broad scenario waiting for further investigation.

In this paper, we theoretically explore the phase diagram and bonding of Xe-Ar binary system under high pressures. Our results demonstrate the existence of a Xe-Ar compound with the stoichiometry of XeAr$_2$. Due to the fact that both of Ar and Xe have filled valence shells, this compound is bound by vdW forces under low pressures. However, high pressures modified substantially the atomic interactions in this compound and strong covalent bonds have been detected.

## 2. COMPUTATIONAL DETAILS

We perform a systematic structural search for the Xe-Ar binary system based on a particle swarm optimization methodology as implemented in the CALYPSO code[40,41], which has been successfully employed in predicting a large variety of crystal structures[9,17-19,42-44]. The underlying total energy calculations and structural relaxations are carried out within the framework of density functional theory using the projector-augmented-wave (PAW) method[45] as implemented in the VASP code[46]. We adopt the Perdew-Burke-Ernzerhof generalized gradient approximation[47] to describe the exchange−correlation functional. The electron-ion interaction is described by pseudopotentials with $5s^25p^6$ and $3s^23p^6$ as valence electrons for Xe and Ar, respectively. The use of a cutoff energy of 650 eV and dense enough k-point sampling grids give excellent convergence of the calculated enthalpy (<1 meV/atom). The dynamical stability of predicted structures is determined by phonon calculations using the finite displacement approach as implemented in the Phonopy code[48].

## 3. RESULTS AND DISCUSSION

In order to obtain the most energetically favorable structures for the Xe-Ar binary system, the stoichiometries of XeAr$_n$ ($n$ = 1-8) containing up to 4 formula units (f.u.) per simulation cell are systematically searched under pressures of 0, 100, 200 and 500 GPa. The calculated formation enthalpies ($H^f$) of each stoichiometry at different pressures are shown in the form of convex hulls as depicted in Figure 1. From this figure, the thermodynamic stability of XeAr$_n$ can be determined. The results indicate that each stoichiometry of XeAr$_n$ has a small positive $H^f$ at 0 GPa. As pressure increase to 100, 200 or 500 GPa, the convex hulls are dominated by a well-developed minimum at $XeAr_2$ stoichiometry, indicating that only $XeAr_2$ is stable against decomposition in to elemental Xe and Ar. The other stoichiometries are thermodynamically unstable due to the fact that the increase of pressure would promote their formation enthalpies to be more positive.

The stable $XeAr_2$ compound is predicted to crystallize in the $MgCu_2$-type Lave phase structure (Figure 2a), wherein the Xe atoms occupy the Mg (8a) sites, forming a diamond-type sublattice, and the Ar atoms reside in the voids of the Xe framework, occupying the Cu (16d) sites. By representing its constituents as hard spheres, the stability of $AB_2$-type Lave phase can be understand by packing rule, i.e., a hard-sphere radius ratio $R_A/R_B$ close to 1.25 will achieve the maximum packing efficiency (0.71%) when crystalize in Lave phase[49]. The ratio $R_A/R_B$ of $XeAr_2$ is 1.13, which is close to the ideal value of a Laves phase. Our structural search simulations reveal that the increase of pressure will further stabilize the Laves phase $XeAr_2$, instead of any phase transition or decomposition, which is in sharp distinction with other NG compounds such as $XeHe_2$[32], $NeHe_2$ [26] and $ArHe_2$[26].

The inset of Figure 1 shows the pressure-dependent formation enthalpy of $MgCu_2$-type $XeAr_2$ wherein it is shown that $XeAr_2$ becomes stable at 1.1 GPa. The effect of vdW interactions on the stability of $XeAr_2$ is also calculated by using the PBE-D2 method[50]. The results indicate that the stable pressure shift to 2.6 GPa with the inclusion of vdW corrections. To determine the dynamical stability of $MgCu_2$-type $XeAr_2$, the phonon dispersion spectra are calculated based on the quasi-harmonic model, and the selected results are shown in Figure 3. It is found that $XeAr_2$ is dynamically stable without showing any imaginary phonon frequency in the pressure range from 5 GPa to 500 GPa.

It is known that the Gibbs free energy (G = H - TS) reduces to enthalpy (H = U + pV) when temperature is 0 K. The formation enthalpy of a compound is determined by the relative internal energy $\Delta U$ and $p\Delta V$ term with respect to elemental solids. The calculated pressure dependence of $\Delta U$ and $p\Delta V$ of $XeAr_2$ are shown in Figure 4. It is obvious that the $p\Delta V$ term has negative values over the pressure range considered. The large gain in $p\Delta V$ term effectively tunes the formation enthalpy to be negative though the relative internal energies $\Delta U$ are positive at 0~150 GPa, leading to the reaction of Xe with Ar above 1.1 GPa. The negative $p\Delta V$ of $XeAr_2$ is attributed to the smaller volume compared to the mixing volume of elemental Xe and Ar. For comparison, we also calculated the pressure dependent $p\Delta V$ for MgCu$_2$-type XeS$_2$[18] and AlB$_2$-type XeHe$_2$[32]. As shown in Figure 4, the $p \sim p\Delta V$ curve of XeS$_2$ is similar to that of XeAr$_2$, being negative in the whole pressure region considered. While the results of XeHe$_2$ is negative under low pressures but becomes to be positive under high pressures. In other worlds, the pressure induced close-packed MgCu$_2$-type XeS$_2$ and XeAr$_2$ make themselves energetically more favorable than the elemental solids in the whole pressure region considered, whereas the AlB$_2$-type XeHe$_2$ which adopts a layered structure would stabilizes at low pressures but decompose at high pressures[32].

To get further insight into the stabilized mechanism, we calculated the electronic properties and atomic interactions of XeAr$_2$ under high pressures.

Figure 5 shows the calculated projected density of states (PDOS) at different pressures. It is clear that XeAr$_2$ is a wide-gap insulator under low pressures. At 10 GPa, the band gap is 8.0 eV (Figure 5a). The Ar 3s and Xe 5s electrons occupy the lowest states of valence band located at -17.5 eV and -12.5 eV, respectively. The states range from -5.5 eV to 0 eV are predominantly composed of Ar 3p and Xe 5p components mixed with a small contribution from Ar 3p and Xe 5p. All of these states are highly localized, which suggest a weak interaction between Xe and Ar. As pressure increases, the band gap decreases slowly, and the metallization occurs at about 500 GPa (Figure 2b). Under high pressure, as can be seen from the PDOS plotted in Figure 5b, the electronic states are wildly broadened, indicating the relevant valence electrons become more delocalized. Moreover, the PDOS shows a significant overlap between the 5p states of Xe and the 3p states of Ar. Visible overlaps also exit between other states, such as the Ar 3s and Xe 5p, and the Xe 5s and Ar 3p states. Additionally, Bader analysis[51] reveals a charge transfer of 0.43 $e$/atom from Xe to Ar at 500 GPa (Figure 2C). These features of electronic properties indicate strong chemical interactions in the XeAr$_2$ compound.

Given the fact that NG atoms have filled valence shells, two NG atoms always do not engage in chemical bonding. However, the pressure induced formation of Xe−Xe covalent bond is predicted in a Xe$_2$F compound, which is the first example of NG-NG bond. For XeAr$_2$ that composed by two NG elements, the nearest Xe-Xe, Xe-Ar and Ar-Ar distances at 500 GPa are 2.56 Å, 2.45 Å and 2.09 Å, respectively. These atomic distances are close enough for them to form covalent bonds compared with the proposed covalent radius of Xe (1.40 Å) and Ar (1.06 Å) atoms[52]. To detect the covalent bonding in XeAr$_2$, we calculate the Crystalline Orbital

Hamiltonian Population (COHP) for nearest-neighboring Xe-Xe, Xe-Ar and Ar-Ar pairs. As can be seen form Figure 5c, the contributions from bonding states are larger than that from antibonding in the occupied states of Xe-Xe pair. As a result, the integrated COHP (ICOHP) up to the Fermi level is -1.20 eV/pair for bonded Xe-Xe, indicating a strong covalent bonding. Furthermore, we also calculate the electron localization function (ELF)[53] of XeAr$_2$ at 500 GPa (Figure 2d). The observed large ELF basins (~0.5) confirm the presence of covalent Xe-Xe bonds. The COHP plots of Xe-Ar and Ar-Ar pairs show that the visible overlaps of the Ar 3s, Ar 3p, Xe 5s and Xe 5p states range from -35 to -15 eV are also related to the covalent bonding. However, because of the large occupation of the antibonding states, the ICOHP of Xe-Ar is -0.36, indicating a weaker covalent bonding. The obtained smaller ELF values (0.32) between the neighboring Xe-Ar atoms and the ELF distributions of Ar atoms depart from sphericity also confirm this. The bonding of neighboring Ar-Ar atoms is much weaker since the ICOHP is only -0.07.

## 4. CONCLUSIONS

In summary, we have theoretically explored the stability of XeAr$_n$ ($n$ = 1-8) under high pressures up to 500 GPa using the effective CALYPSO structure searching method combined with first-principles calculations. The results shown that Xe reacts with Ar at a pressure as low as 1.1 GPa, producing a MgCu$_2$-type Lave phase XeAr$_2$ compound. This compound is a wide-gap insulator which would metallizes at about 500 GPa. Analysis of the electronic structures and atomic interactions indicates the vdW XeAr$_2$ compound under low pressures would change to an extended solid wherein atoms are bonded by network of covalent bonds under high pressures. Our results shed light on the chemical bonding between NG elements and enrich NG chemistry.

**ACKNOWLEDGMENTS**

This work is supported by the National Natural Science Foundation of China under Grant Nos. 11672274 and 11274281, the CAEP Research Projects under Grant Nos. 2012A0101001 and 2015B0101005, the NSAF under Grant No. U1730248, and the Fund of National Key Laboratory of Shock Wave and Detonation Physics of China under Grant No. 6142A03010101.

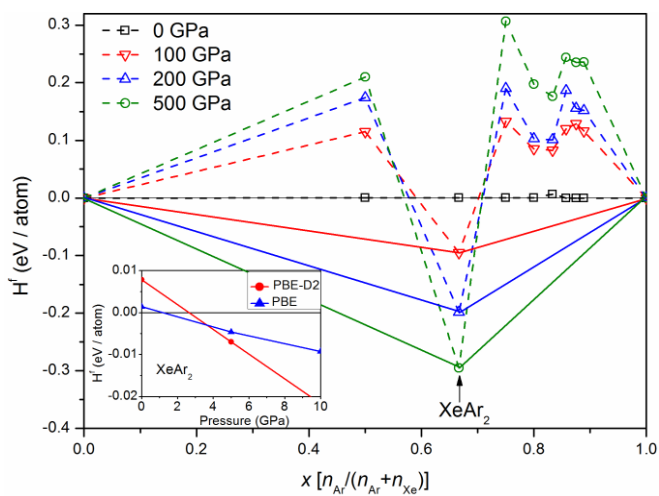

Figure 1. Formation enthalpies ($H^f$) of XeAr$_n$ ($n$ = 1-8) with respect to decomposition into constituent elemental solids under different pressures. Solid lines denote the convex hull, where the data points located on represent stable species against any type of decomposition. Inset: Pressure-dependent formation enthalpy of XeAr$_2$ obtained with (PBE + D2) and without (PBE) vdW corrections.

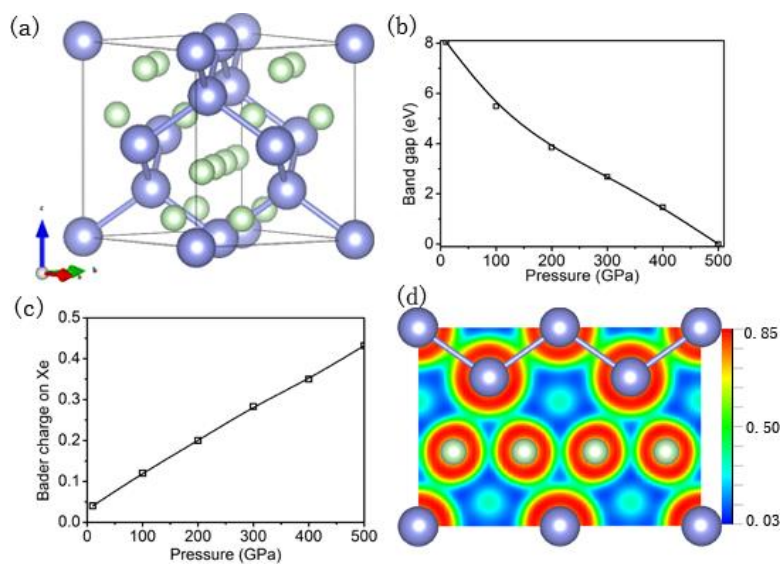

Figure 2. (a) Crystal structure of XeAr$_2$. The big and small spheres represent Xe and Ar, respectively. (b) Pressure-dependent band gap of XeAr$_2$. (c) Pressure-dependent Bader charge on the Xe atoms in XeAr$_2$. (d) Electron localization function (ELF) of XeAr$_2$ at 500 GPa.

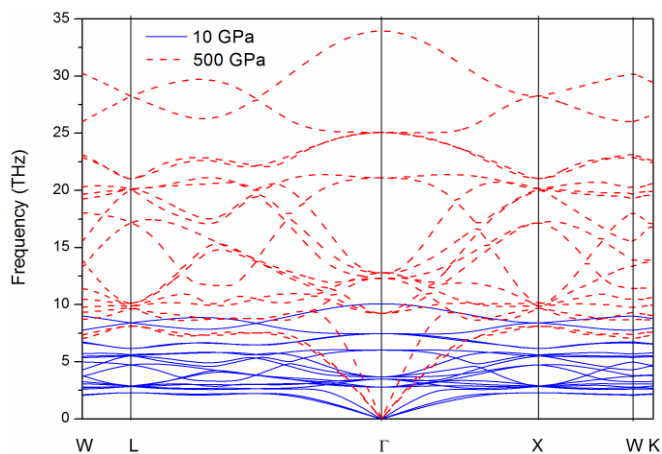

Figure 3. Phonon dispersion spectrum of XeAr$_2$ under different pressures.

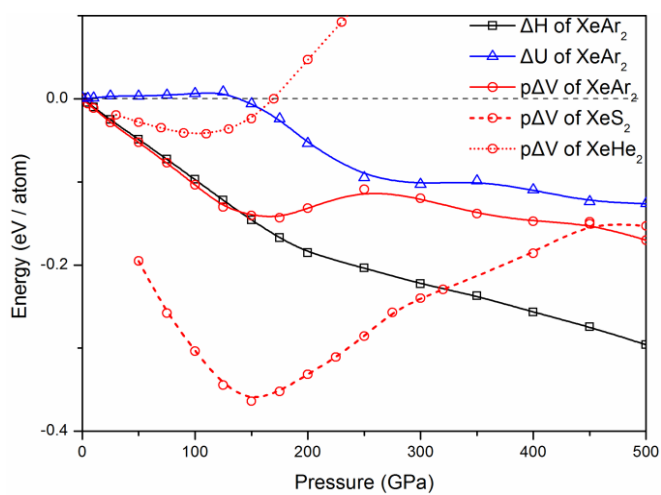

Figure 4. Pressure dependence of $\Delta H$, $\Delta U$, and $p\Delta V$ of relevant compounds.

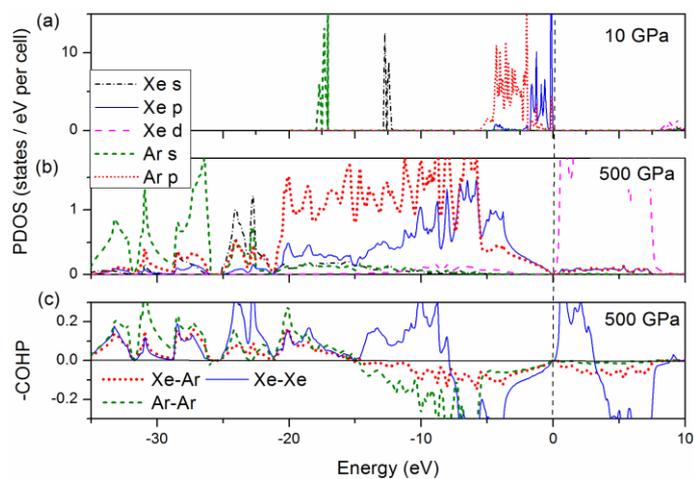

Figure 5. (a, b) Projected density of states (PDOS) of XeAr$_2$ at different pressures. (c) Crystal orbital Hamilton population (COHP) plots for nearest-neighboring Xe-Ar, Xe-Xe and Ar-Ar pairs at 500 GPa.